\documentclass[conference]{IEEEtran}

\usepackage[tight,footnotesize]{subfigure}
\usepackage{graphicx,epstopdf} 
\usepackage{cite}
\usepackage[pdftex]{hyperref}
\hypersetup{
pdftitle={Your Title}, pdfauthor={Your Authors}, pdfkeywords={your
keywords}, bookmarksnumbered, pdfstartview={FitH}, colorlinks,
citecolor=black, filecolor=black, linkcolor=black, urlcolor=black,
breaklinks=true, }

\begin{document}
\title{Unconventional TV Detection using Mobile Devices}

\author{\IEEEauthorblockN{Mohamed Ibrahim, Ahmed Saeed and Moustafa Youssef}
\IEEEauthorblockA{Department of Computer Science and Engineering\\
Egypt-Japan University of Science and Technology (E-JUST), Egypt\\
Email: \{mibrahim,ahmed.saeed,moustafa.youssef\}@ejust.edu.eg}
\and \IEEEauthorblockN{Khaled A. Harras}
\IEEEauthorblockA{Computer Science Department\\
School of Computer Science\\
Carnegie Mellon University Qatar \\
Email: kharras@cs.cmu.edu }}

\maketitle

\begin{abstract}
Recent studies show that the TV viewing experience
is changing giving the rise of trends like
\emph{"multi-screen viewing"} and \emph{"connected viewers"}.
These trends describe TV viewers that use mobile devices
(e.g. tablets and smart phones) while watching TV.
In this paper, we exploit the
context information available from the ubiquitous mobile devices to detect the presence
of TVs and track the media being viewed. Our
approach leverages the array of sensors available
in modern mobile devices, e.g. cameras and microphones,
to detect the location of TV sets, their
state (ON or OFF), and the channels they are currently
tuned to. We present the feasibility of the proposed
sensing technique using our implementation on
Android phones with different realistic scenarios.
Our results show that in a controlled environment
a detection accuracy of $0.978$ F-measure could be
achieved.

\end{abstract}


\section{Introduction}
TV viewers' profiling is an important functionality for both advertisers and service providers. Traditionally, the detection techniques of TV viewers' habits are
concerned more about the collective preferences of the viewers
and rely mainly on focus groups \cite{AudienceGroup}
or special hardware connected to the TV (e.g. set top devices)
\cite{thomas1992television}.
Recent studies show that 52\% of cell phone owners use their phones
while watching TV \cite{report1} and 63\% of tablets
owners use their tablets while watching TV \cite{report2}
in what was called \emph{``Connected Viewers''}.
The rise of these \emph{``Connected Viewers''} opens the door for a new unconventional approach for TV viewers' profiling based on the ubiquitous mobile devices and their equipped sensors. Such approach can provide ubiquitous fine-grained information about the user's TV viewing preferences leading to new possibilities for advertisers and service providers on both the TV and mobile sides.
These possibilities include fine-grained audience measurement, tracking mobile users' preferences
through their TV viewing habits, targeted mobile ads,
and combined mobile-TV experience personalization.

Earlier work for TV set detection, e.g. \cite{DetectPR}, relied on
special devices that can detect the power leakage of a TV receiver's local
oscillator. Such systems do not scale and are harder to deploy.
From a different perspective, extensive work has been done in detecting TV shows and commercials \cite{Com}.
This involves 
scene boundary detection \cite{Scence} and TV shows recognition \cite{Social}.
Another line of work depends on audio as their data source for TV shows and music identification including commercial products
\cite{AudioIden, Shazam, ContactMusic,Background}. However,
all these audio detection approaches focus on identifying the content regardless of its source and hence cannot determine the audio source type (whether it is a laptop, people talking, or TV).

In this paper, we present the design, implementation and evaluation
of a system that can leverage the large array of sensors currently available
in smart phones and other mobile devices to accurately
detect the presence of TV sets. In particular, our implementation currently
depends on the acoustic context analysis and visual surroundings detection using
microphones and cameras embedded in mobile devices to identify (1) the presence and
locations of the TV sets, (2) whether they are ON or OFF, and (3) the
channels they are currently tuned to.
We test the system's ability to differentiate
between a TV set's acoustic and visual fingerprints
on one side and other sources of similar fingerprints
such as people having a conversation and laptops playing audio/video
files on another side. The results showed that a typical mobile device
can reach an F-measure of $0.978$ in a \emph{controlled environment}.

\section{System Architecture}

\begin{figure}[!t]
\centering
  \includegraphics[width=0.4\textwidth]{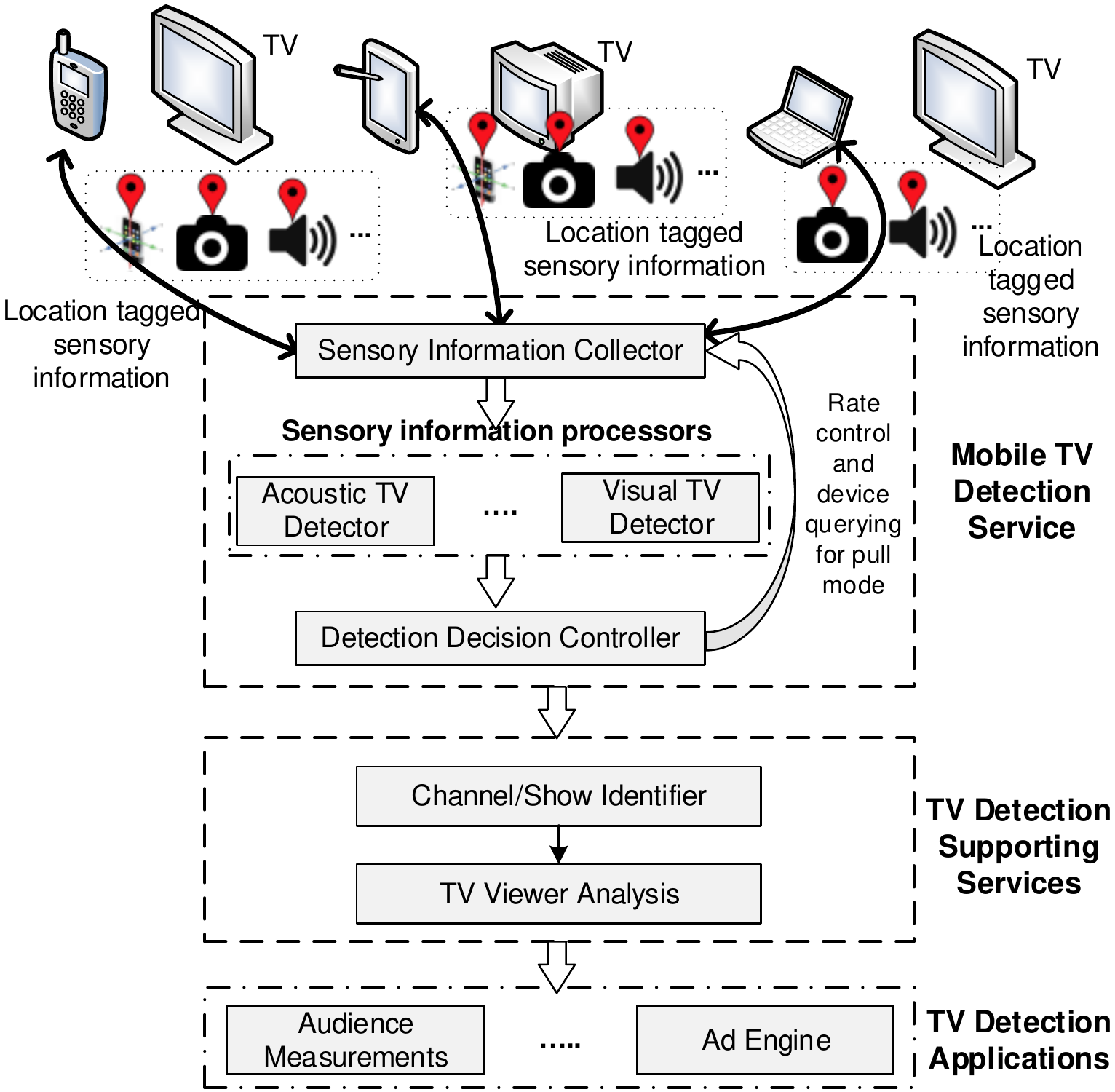}
  \caption{System architecture.}
  \label{fig:system}
\end{figure}

Figure~\ref{fig:system} shows the proposed architecture.
Standard mobile devices with embedded sensors
(such as mic, camera, accelerometer, GPS, etc)
submit their location tagged sensory information to
the system's server. The server has three main
components:  Mobile TV Detection Service,
TV Detection Supporting Services and
TV Detection Applications.

\textbf{Mobile TV Detection Service} is responsible
for the collection and processing of the sensory information.
This service contains different modules responsible
for the detection of TV sets based on information collected
from different sensors. It is also responsible for the fusion
of the detection decision made by the different sensors. Moreover,
this service is responsible for controlling the rate at
which the sensors collect their information.

\textbf{TV Detection Supporting Service} is responsible
for further processing of the information collected about
the detected TV sets. It connects to TV streaming servers,
online schedules, and media databases to detect the current channel.
It depends on the comprehensive previous techniques for detecting
TV shows, music, and commercials \cite{AudioIden, Com, Shazam}. Other possibilities include interaction with social 
network sites to access information about the user preferences.

\textbf{TV Detection Applications} use the TV sets
information collected by other services to provide different
services either to the mobile user (e.g. personalization)
or to third party applications (e.g. audience measurement
systems and targeted ads systems).

For the rest of the paper, we focus on the detection of
the presence of a TV sets using mobile phones.
We present the design, implementation and evaluation
of the \emph{Mobile TV Detection Service}.

\section{Mobile TV Detection Service}

We implemented the service on different Android
phones and used it while watching different TV
sets made by different manufacturers. We tested
our implementation in a \textbf{controlled environment}
using two sensors: microphone and camera. We address the challenge of
differentiating the visual and acoustic signature of TV
sets and other sources that could be confused with the TV.
For example, the sounds coming from a TV set could be confused
with a laptop or normal people talking. Moreover, the
visual signature of a TV set (i.e. rectangular-shaped object with
varying content) could be confused with picture frames and
windows.

\subsection{Acoustic TV Detector}

\begin{figure}[!t]
\centering
     \subfigure[Movie on a TV and laptop]{\includegraphics[width=0.22\textwidth]{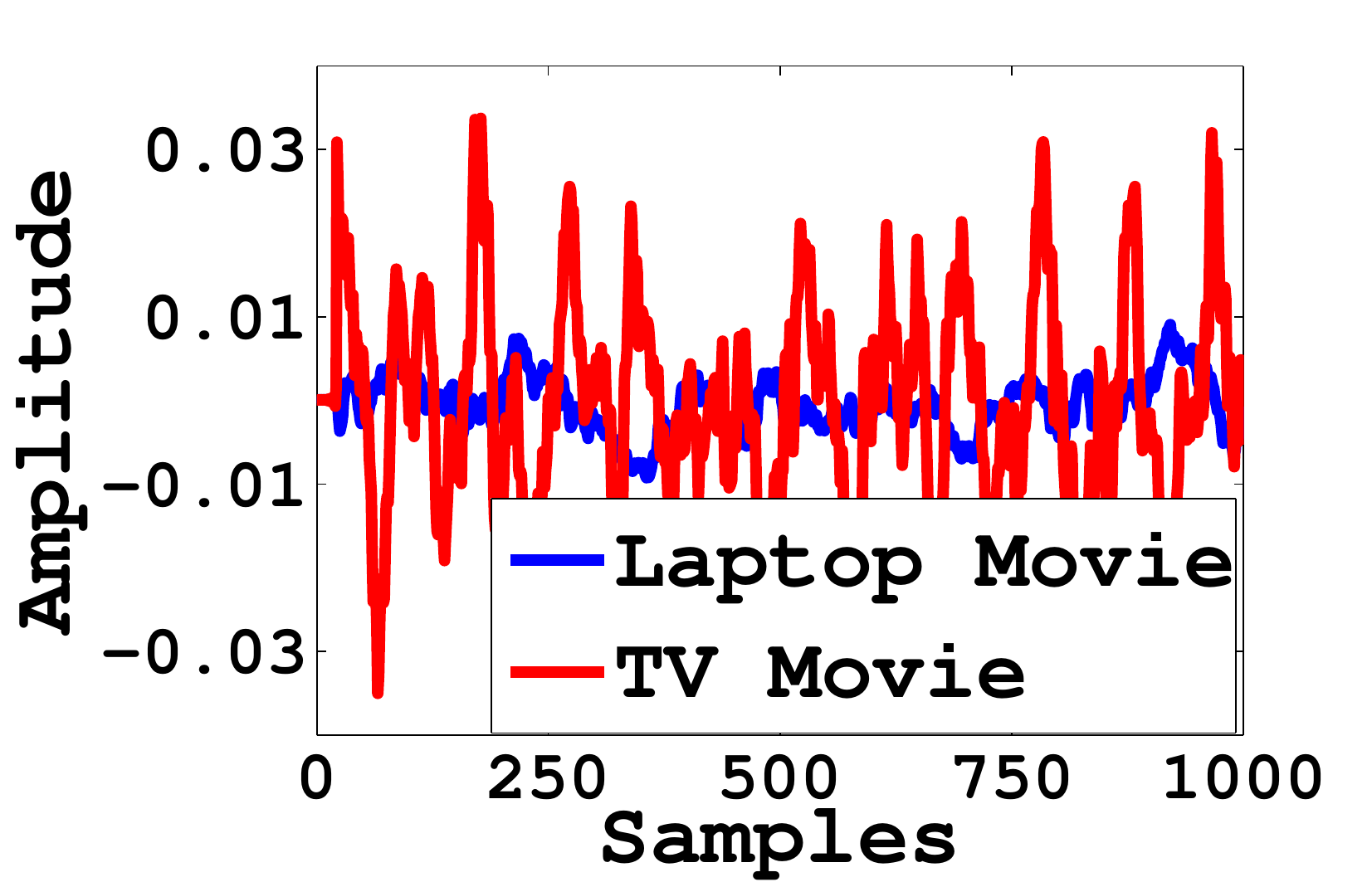}\label{fig:rawMicLaptop}}
     \hfill
     \subfigure[TV show and normal people talking]{\includegraphics[width=0.23\textwidth]{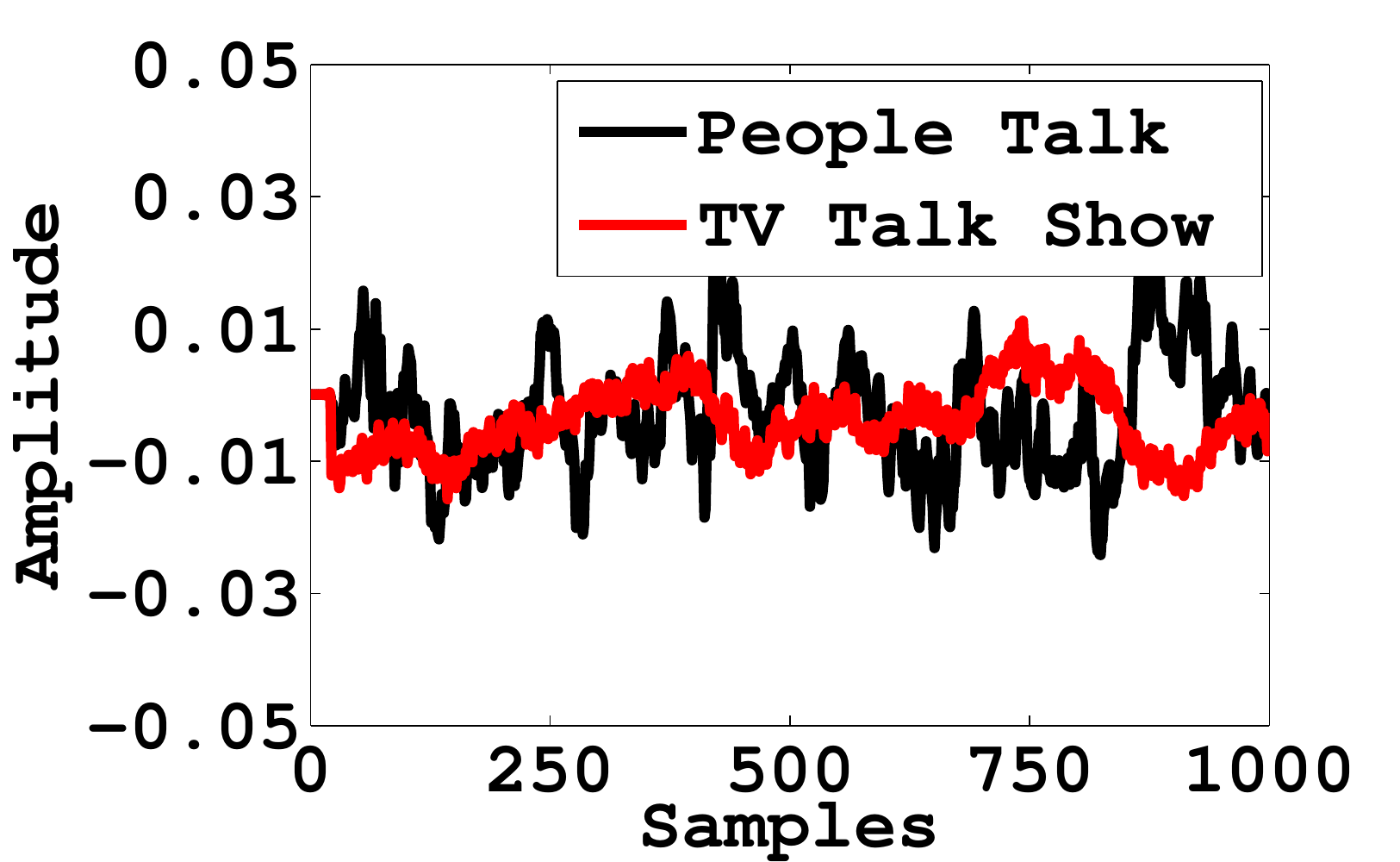}\label{fig:rawMicTalking}}
     \hfill
      \caption{Acoustic time-domain raw data amplitude.}     \label{fig:raw_data}
 \end{figure}

\begin{figure}
\centering
     \subfigure[Watching a movie on laptop vs. TV]{\includegraphics[width=0.23\textwidth]{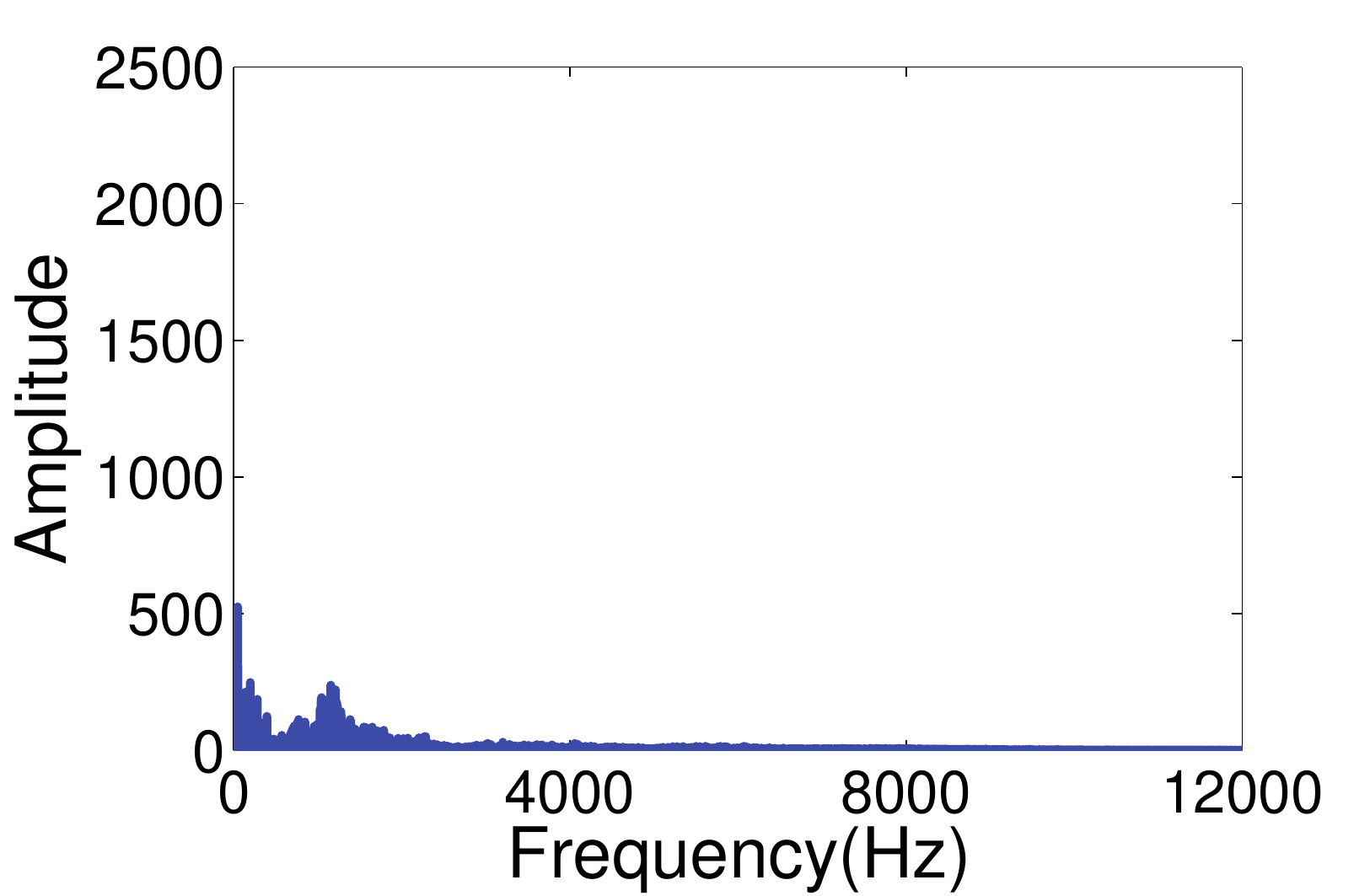}\label{fig:rawFreqLaptop1}
     \includegraphics[width=0.23\textwidth]{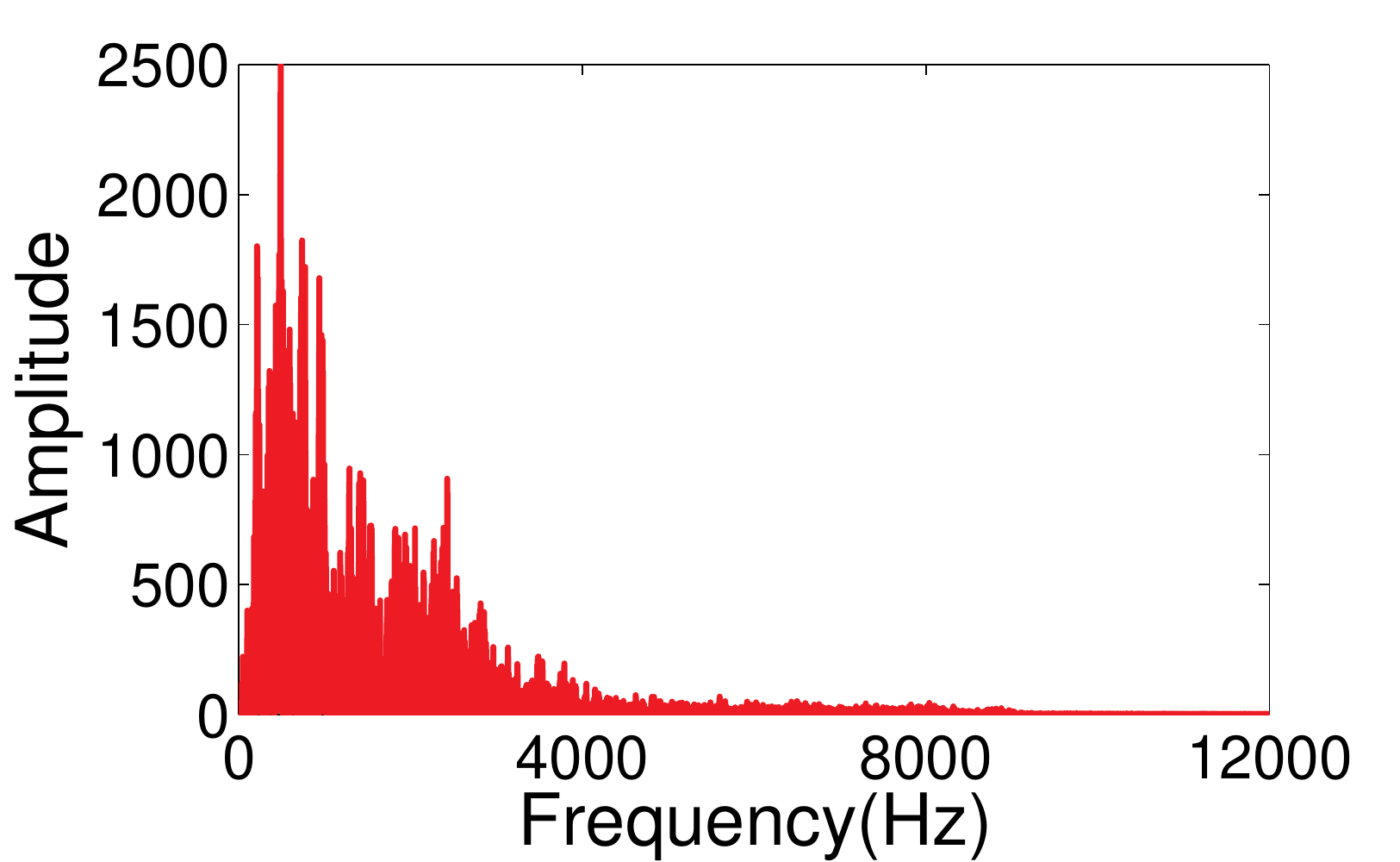}\label{fig:rawFreqTV}
     }
     \vfill
     \vspace{-0.15in}
     \subfigure[Normal people talk vs. Watching a talk show on TV ]{\includegraphics[width=0.23\textwidth]{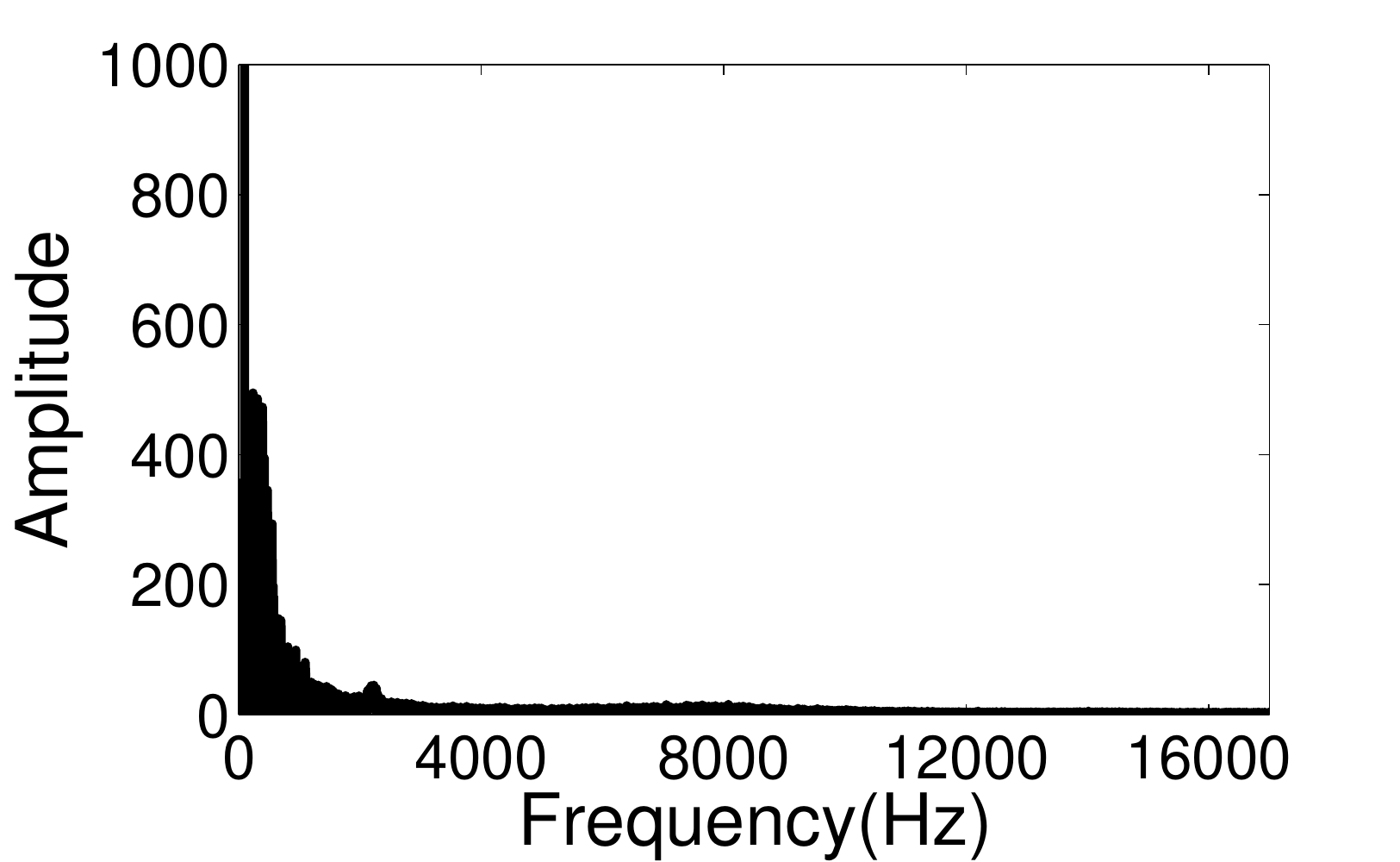}\label{fig:rawFreqLaptop2}
     \includegraphics[width=0.23\textwidth]{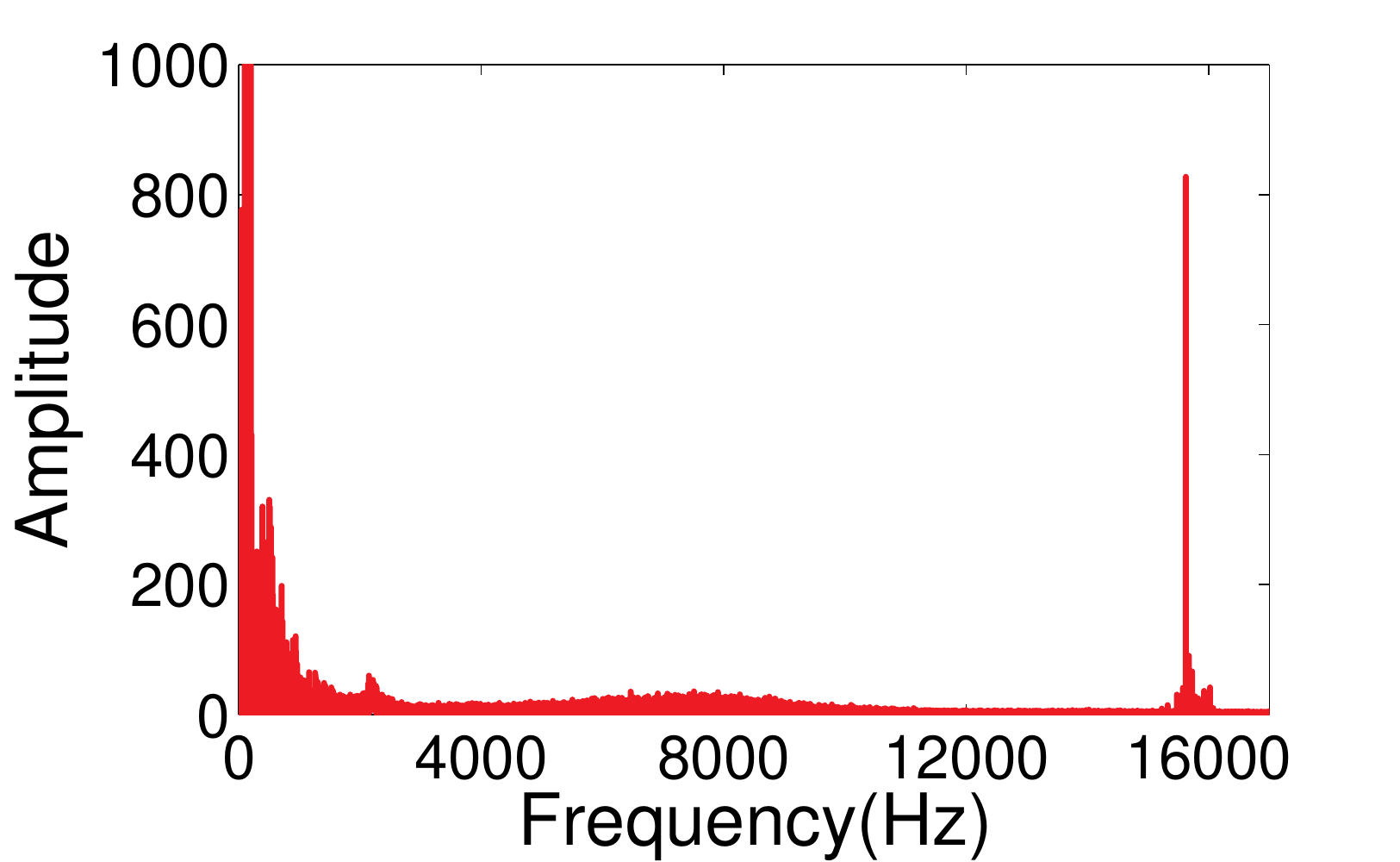}\label{fig:rawFreqTVTalks}
     }

      \caption{Acoustic frequency-domain raw data.}     \label{fig:freq_domain}
 \end{figure}

The main challenge for acoustic TV detection is extracting unique
features for the acoustic fingerprint that would enable the differentiation
between TV sets and other sources that could be confused with it.
We collected an acoustic dataset composed of $91$ audio recordings
for training and $60$ independent audio recordings for testing.
Each audio recording is $30$ seconds long. We had different configurations
including the TV volume, phone relative distance to the TV, position of
the phone (in pocket, on couch, etc), show type (movie, talk show, sports,
etc), gender and talking level of the actor/anchor. Also, we collected a
data set under the same different configurations for the laptop and
normal people talk classes.
Our goal in the rest of this section is to identify time and frequency domain features
that can differentiate between the TV case on one hand and
\{the laptop and people talking\} case on the other hand.
Figures \ref{fig:raw_data} and \ref{fig:freq_domain} show sample raw
data obtained from our acoustic dataset.

\subsubsection{Time domain features}
Figure~\ref{fig:rawMicLaptop} shows the raw time domain
acoustic amplitude for listening to a movie on a TV and on a
laptop whereas Figure \ref{fig:rawMicTalking} shows the same signal
while listening to a TV show and listening to a group of people
talk. The figure shows that there is a noticeable difference between
the two cases in each figure. This is intuitive as a person listening
to a movie or show on a laptop will usually have a lower volume than
the case of listening to the same show on the TV. On the other hand,
people talking will tend to lower the volume of the TV.

Based on Figure~\ref{fig:raw_data}, we extract features that capture
the amplitude variations of the acoustic signal in the
time domain. One of these key features
is the Zero Crossing Rate (ZCR) \cite{Features1} that represents the rate
at which the signal crosses the x-axis (zero amplitude).
ZCR is used to estimate the fundamental frequency of the
signal. Therefore, it is used as an indication of the noisiness of the signal.
Another time domain feature is the Short Time Energy (STE) \cite{Features1}
that captures the loudness of the signal and is computed as
the average of the square amplitude of the signal.

\subsubsection{Frequency domain features}
Figure~\ref{fig:freq_domain} shows the frequency domain signal
for the same example in Figure~\ref{fig:raw_data}. The figure
shows that the frequency domain response of the signal differs
from the TV and other classes. From the figure, it could be observed
that media streamed to laptops are lower quality in terms of bit rate
compared to media displayed on the TV. This observation leads
to the conclusion that the acoustic fingerprint of laptops
will have a lower bandwidth as compared to TV sets. Similarly,
comparing the acoustic fingerprint of a TV set and normal people
talk, it could be observed that the TV set's fingerprint
is a combination of people talk (4 KHz) and other sources
(e.g, music (16 KHz)). This observation also leads to the conclusion
that people conversations will have a lower bandwidth
as compared to TV sets in the frequency domain. Based on these
observation, we use the following frequency domain features:
Spectral Centroid (SC) and Spectrum Spread (BW) \cite{Features1}.
These features represent the spectrum by its center of
mass and its spread around that center.
We also use the Mel-frequency Cepstral Coefficients (MFCC) \cite{Features1}
which are a set of features, where each feature represents
a portion of the spectrum in the Mel scale.

\subsubsection{Acoustic fingerprint classification}
\begin{figure} [!t]
\centering
      \includegraphics[width=0.3\textwidth]{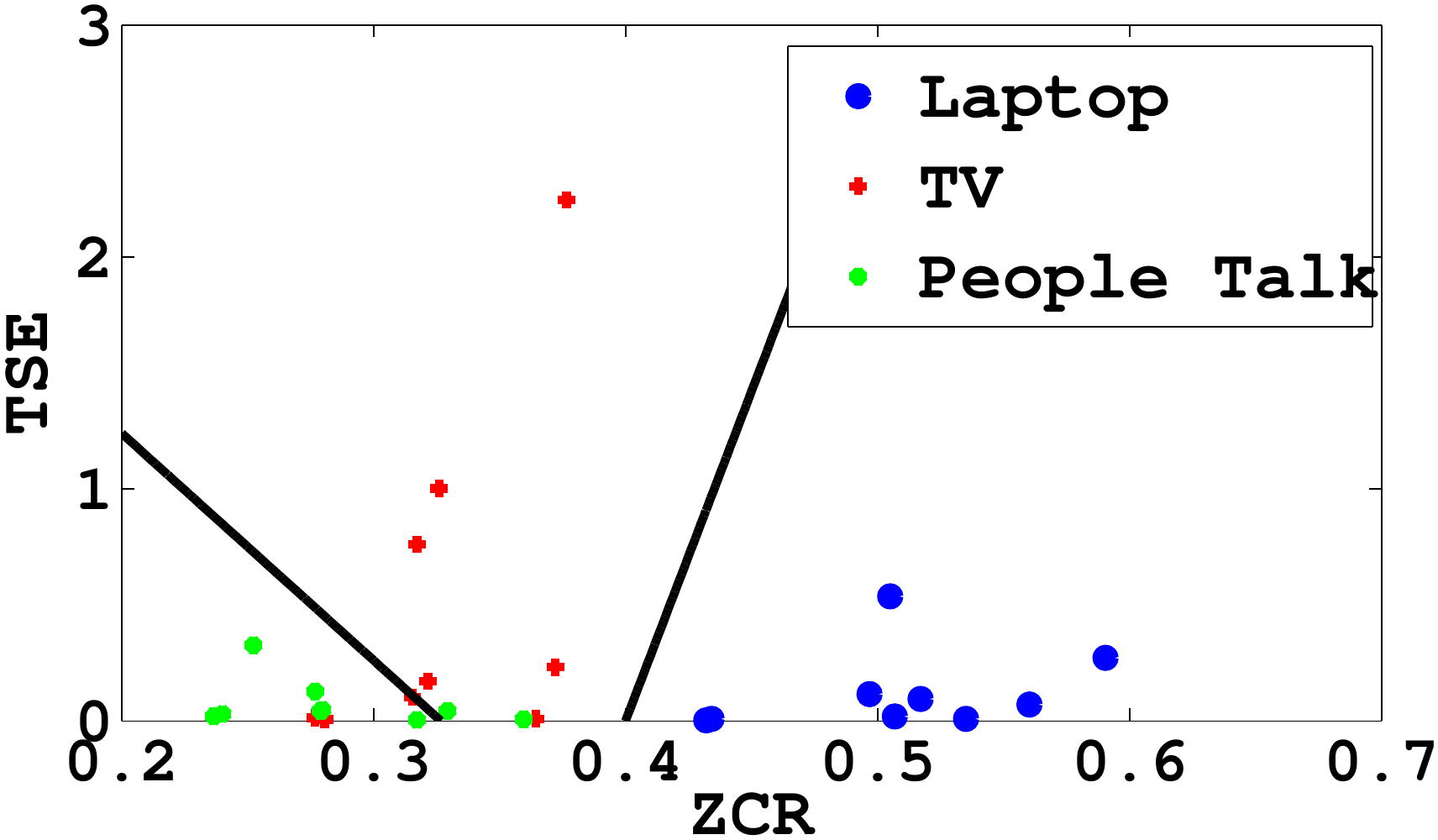}
  \caption{SVM discriminant function using two features.}
  \label{fig:svm}
\end{figure}

After extracting the features, we use a Support Vector Machine classifier to distinguish TVs from the other two classes.
Figure \ref{fig:svm} shows a sample result using the classifier for two
features (ZCR and STE). 

\subsection{Visual TV Detector}

\begin{figure*}[!t]
  \centering
  \subfigure[{\scriptsize The changing areas between different frames are detected as foreground.}]{\includegraphics[width=0.3\textwidth]{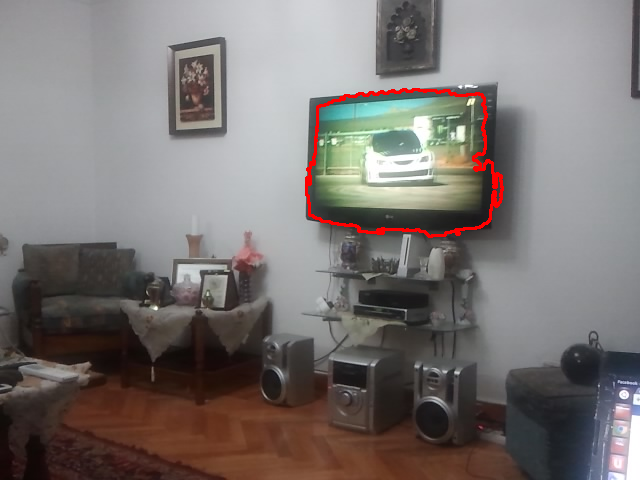}\label{fig:camera1}}
  \subfigure[{\scriptsize All rectangles in picture are detected then small rectangles and large rectangles are filtered out.}]{\includegraphics[width=0.18\textwidth]{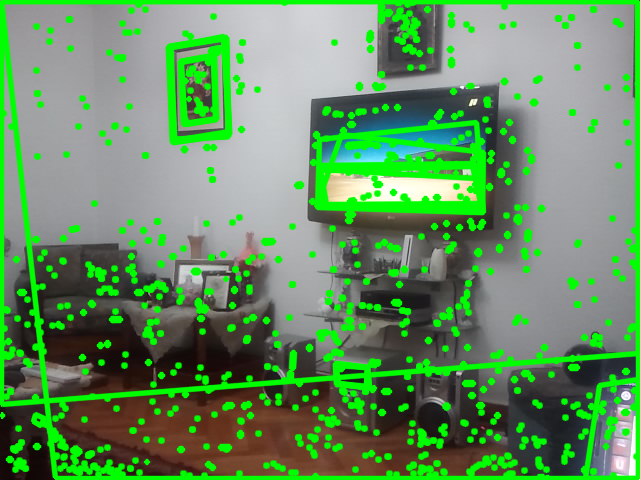}
  \includegraphics[width=0.18\textwidth]{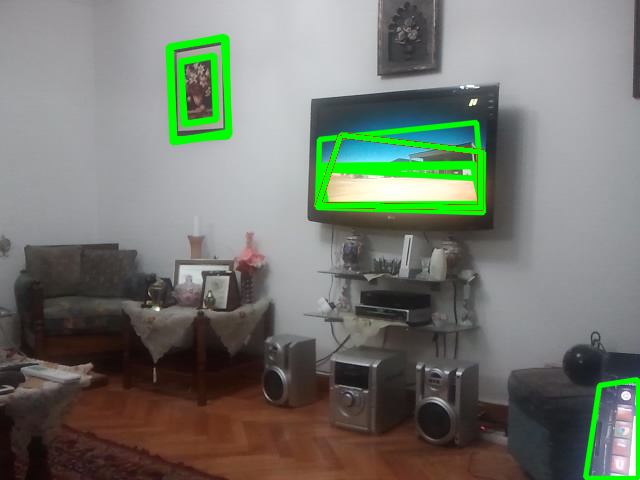}\label{fig:camera2}}
  \subfigure[{\scriptsize The intersection of the previous two steps is performed to detect the existence of the TV.}]{\includegraphics[width=0.3\textwidth]{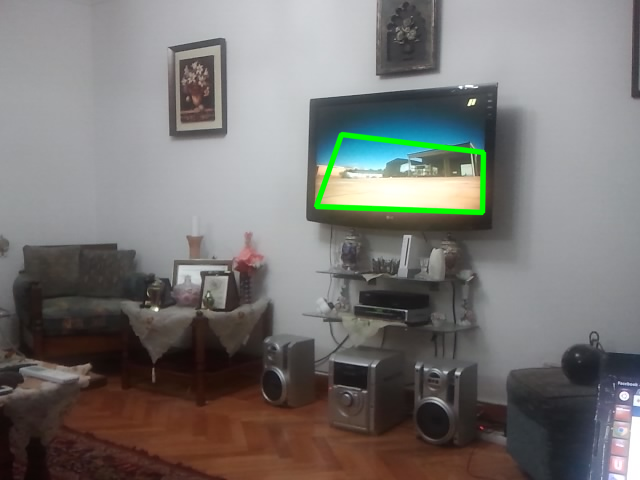}\label{fig:camera3}}
  \caption{TV detection steps using the camera.}
  \label{fig:Camera}
\end{figure*}
Acoustic detection may confuse the presence
of TV sets with other sources of similar acoustic fingerprints,
e.g. stereo players. To reduce this ambiguity, we consider
the usage of cameras as a source of detection information.
Our approach is based on recent statistics that
show that a smart phone user holds the smart phone
at $12.5$ inches from her eyes while surfing and at $14.1$ inches away from her eyes while texting
\cite{bababekova2011font}. At these distances, if the smart phone user is watching the TV, the TV will either
 partially or fully appear within the camera's frame. We collected $26$ shots by normal users using their phones,
e.g. to text or browse the Internet, with each shot composed of $8$ consecutive
frames. Fourteen out of the 26 shots were taken in different light conditions
in different locations with a mix of shots showing the TV as a whole or partially.
The remaining 12 shots had no TV sets but rather objects that could be confused with TV sets
using our proposed algorithm (e.g. windows, doors, picture frames, and shots with moving people and objects).

We use a simple algorithm that detects the characteristics
of a TV in a sequence of frames captured either through a recorded video or sequence of captured images.
The algorithm works in three steps summarized in Figure \ref{fig:Camera}.
The first step, Figure \ref{fig:camera1}, detects changing parts in the image sequence, which represent the dynamics of the scene on a TV set. This is performed using a simple background extraction algorithm \cite{zivkovic2004improved}.
In the second step, Figure \ref{fig:camera2}, it determines rectangle-shaped contours within each
image filtering out small contours (smaller than 5\% of the total area of
the image) and large contours (larger than 70\% of the total area of the
image). The rectangle shapes detection works in two steps: the first finds
all contours in the image using the algorithm proposed in \cite{suzuki1985topological}.
In the second step, all contours are simplified by reducing the number of points forming them
using the Ramer-Douglas-Peucker algorithm \cite{douglas1973algorithms}. Reduced convex contours composed of only four points are selected as rectangle-shaped areas.
Finally an intersection step between the previous two stages are performed to produce the final output (Figure \ref{fig:camera3}). In particular, the
rectangle with the smallest area that encloses
all the recorded foreground contour centers is declared to be a TV.

\begin{figure} [!t]
\centering

\subfigure[{\scriptsize Audio sampling rate effect on detection accuracy.}]{\includegraphics[width=0.23\textwidth]{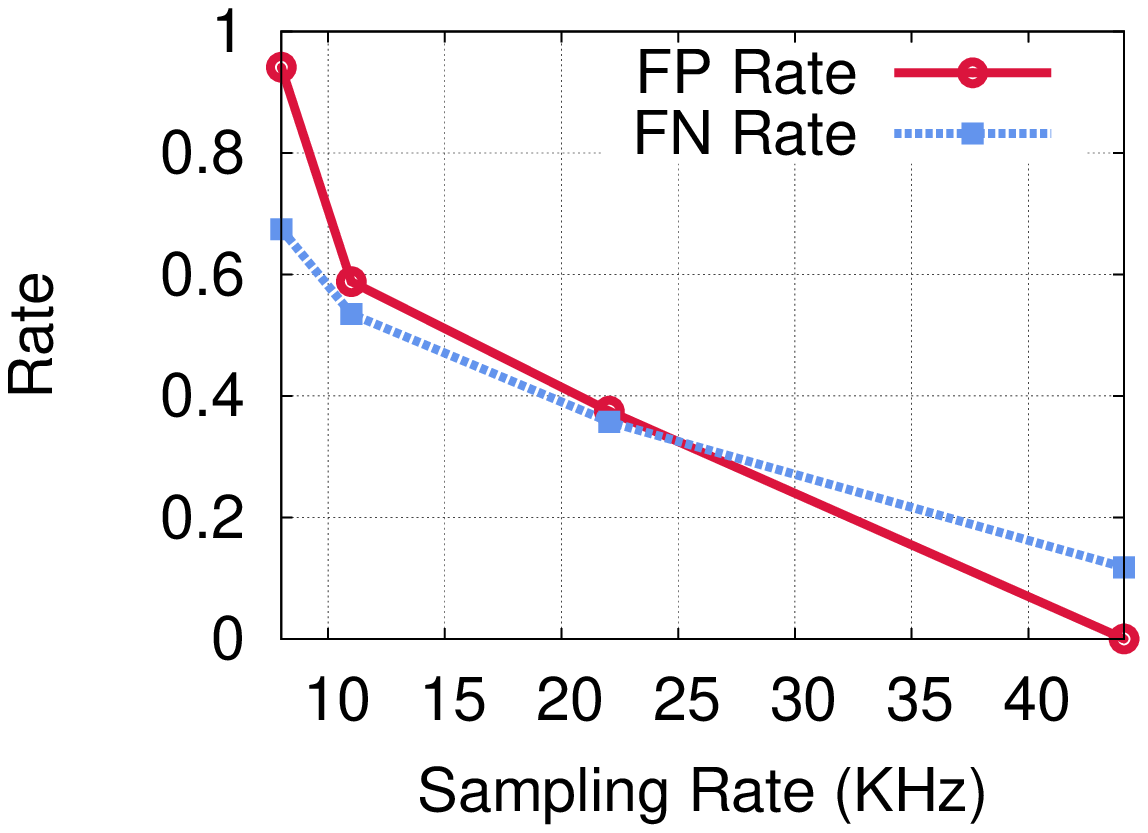}\label{fig:MIC}}
\subfigure[{\scriptsize Camera frames number effect on false positive and false negative rates.}]{\includegraphics[width=0.21\textwidth]{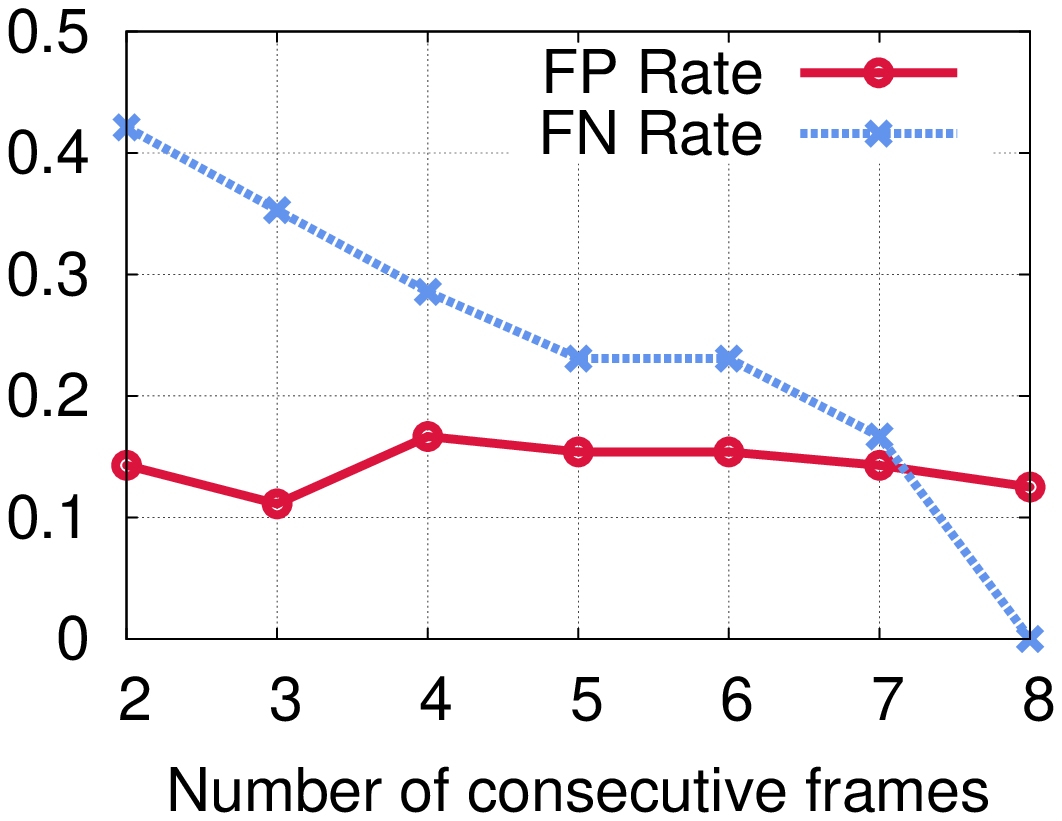}\label{fig:camera_results}}
\caption{Summary of Individual Sensor Results}
\end{figure}

\begin{table}[!t]
\small
\centering
\begin{tabular}{|l||l|l||l|}
\hline

Approaches & Acoustic& Visual& \textbf{Fused}\\
\hline \hline False Negative Rate& 0.13 & 0&\textbf{ 0}\\
\hline False Positive Rate & 0&0.125& \textbf{0.042}\\
\hline F-measure & 0.928 & 0.933& \textbf{0.978}\\

\hline

\end{tabular}
\caption
{Comparison between different TV detection approaches.}
\label{compare}
\end{table}

\subsection{Detection Decision Controller}

This module is responsible for fusing the decisions made
based on different sensors. Furthermore, it is also responsible
for setting the frequency by which the sensory information
are collected (e.g. acoustic sampling rate and number of captured frames
per second) to control both the accuracy and energy
efficiency of the system. The sensors fusion is based on the assumption that avoiding false negatives is more 
important than false positives as not
detecting a TV set wastes opportunities of detecting context
information. Therefore, we fuse
the results of the audio and video modules using the
a simple OR rule: If the two techniques result in two
opposite results, then the fused results will be always positive,
i.e a TV is detected.

Figure~\ref{fig:MIC} shows the effect of the acoustic sampling rate on the false
positive and false negative rates. Figure~\ref{fig:camera_results} shows the effect
of increasing the number of consecutive frames on the visual detection algorithm.
Table~\ref{compare} summarizes the results. The acoustic approach achieves a zero
false positive rate and a $0.13$ false negative rate ($0.928$ F-measure) with most
of the errors in mixing a quite talk show on the TV with normal people talking.
On the other hand, the visual detection approach achieves a detection accuracy of
zero false negative rate and a $0.125$ false positive rate ($0.933$ F-measure).
The results of the fusion approach is summarized in Table~\ref{compare}. This
approach results in a zero false negative rate and $0.042$ false positive
rate ($0.978$ F-measure). Note that this can also be further enhanced by
combining the detection results from different nearby devices and other sensors.

\section{Discussion and Future Work}
\subsection{User Privacy}
Protecting the user privacy can be achieved by local processing of the raw data on the user mobile device and forwarding only the anonymized detection results. This can be extended by forwarding the data from the mobile phone to a more powerful device, such as the user laptop for processing before forwarding to the back end server.

\subsection{Incentives}
To encourage the users to deploy the proposed system, different incentive techniques can be used including providing coupons, recommendation services, among other traditional incentive systems.

\subsection{Using Other Sensors}
The proposed approach can be extended to use other sensors. For example, the inertial sensors (e.g. accelerometer. gyroscope and compass) can be used to better trigger
the acoustic and visual detection sensors based on the detected user activity. Other sensors, such as WiFi, can be used to obtain the device location indoors and hence provide better context information about the user actions and the TV location. 
\subsection{Energy Efficiency}
Continuous sensing on a mobile device can quickly drain the scarce battery resource. Automatically setting the sensing rate and which devices to sense based on their remaining battery, the device context and location, and required information are different steps to address this issue. This is one of the main functionalities of the \textit{Detection Decision Controller}. In addition, offloading the computations to a more powerful user device can also help alleviate this concern.

%
%

%


\small
\bibliographystyle{abbrv}
\bibliography{tv_ubicomp} 

\begin{thebibliography}{10}

\bibitem{report2}
{T}ablets and {M}ulti-{T}asking.
\newblock {\em The {G}f{K} {MRI} i{PANEL} Reporter}, 2012.

\bibitem{Com}
A.~Albiol, M.~Ch, F.~Albiol, and L.~Torres.
\newblock Detection of {TV} commercials.
\newblock In {\em Acoustics, Speech, and Signal Processing, 2004. Proceedings.
  (ICASSP '04). IEEE International Conference on}, volume~3, may 2004.

\bibitem{bababekova2011font}
Y.~Bababekova, M.~Rosenfield, J.~Hue, and R.~Huang.
\newblock Font size and viewing distance of handheld smart phones.
\newblock {\em Optometry \& Vision Science}, 88(7):795--797, 2011.

\bibitem{ContactMusic}
M.~Casey, R.~Veltkamp, M.~Goto, M.~Leman, C.~Rhodes, and M.~Slaney.
\newblock Content-based music information retrieval: Current directions and
  future challenges.
\newblock {\em Proceedings of the IEEE}, 96(4):668 --696, april 2008.

\bibitem{Background}
C.-Y. Chiu, D.~Bountouridis, J.-C. Wang, and H.-M. Wang.
\newblock Background music identification through content filtering and
  min-hash matching.
\newblock In {\em ICASSP 2010}, pages 2414 --2417, 2010.

\bibitem{douglas1973algorithms}
D.~Douglas and T.~Peucker.
\newblock Algorithms for the reduction of the number of points required to
  represent a digitized line or its caricature.
\newblock {\em Cartographica}, 10(2):112--122, 1973.

\bibitem{Social}
M.~Fink, M.~Covell, and S.~Baluja.
\newblock Social- and interactive-television applications based on real-time
  ambient-audio identification.
\newblock In {\em EuroITV}, 2006.

\bibitem{Features1}
G.~Peeters.
\newblock {A large set of audio features for sound description (similarity and
  classification) in the CUIDADO project}.
\newblock Technical report, 2004.

\bibitem{AudioIden}
M.~Ramona and G.~Peeters.
\newblock Audio identification based on spectral modeling of bark-bands energy
  and synchronization through onset detection.
\newblock In {\em ICASSP 2011}, pages 477 --480, may 2011.

\bibitem{Scence}
Z.~Rasheed and M.~Shah.
\newblock {Scene detection in Hollywood movies and TV shows}.
\newblock In {\em Computer Vision and Pattern Recognition, 2003. Proceedings.
  2003 IEEE Computer Society Conference on}, volume~2, pages II -- 343--8
  vol.2, 2003.

\bibitem{AudienceGroup}
C.~A. Russell and C.~P. Puto.
\newblock Rethinking television audience measures: an exploration into the
  construct of audience connectedness.
\newblock {\em Marketing Letters}, 10:393--407.

\bibitem{report1}
A.~Smith and J.~L. Boyles.
\newblock {T}he rise of the “{C}onnected {V}iewer”.
\newblock {\em Pew Internet \& American Life Project}, 2012.

\bibitem{suzuki1985topological}
S.~Suzuki et~al.
\newblock Topological structural analysis of digitized binary images by border
  following.
\newblock {\em Computer Vision, Graphics, and Image Processing}, 30(1):32--46,
  1985.

\bibitem{thomas1992television}
W.~L. Thomas.
\newblock Television audience research technology, today's systems and
  tomorrow's challenges.
\newblock {\em Consumer Electronics, IEEE Transactions on}, 1992.

\bibitem{Shazam}
A.~L. Wang.
\newblock An industrial-strength audio search algorithm.
\newblock In {\em ISMIR 2003}, pages 7--13. ISMIR, 2003.

\bibitem{DetectPR}
B.~Wild and K.~Ramchandran.
\newblock Detecting primary receivers for cognitive radio applications.
\newblock In {\em DySPAN 2005}, pages 124 --130, 2005.

\bibitem{zivkovic2004improved}
Z.~Zivkovic.
\newblock Improved adaptive gaussian mixture model for background subtraction.
\newblock In {\em ICPR 2004}, volume~2, pages 28--31. IEEE, 2004.

\end{thebibliography}

\end{document}